\documentclass[10pt,a4paper]{article}
\usepackage[latin1]{inputenc}
\usepackage{amsmath}
\usepackage{amsfonts}
\usepackage{amssymb}
\usepackage{graphicx}
\begin{document}

\title{Parton discrimination using jets with ALICE at the LHC\footnote{Preprint of an article submitted for consideration in IJMPE \copyright 2010 [copyright World Scientific Publishing Company] http://www.worldscinet.com/ijmpe/ijmpe.shtml}}

\author{HERMES LE\'ON VARGAS for the ALICE Collaboration \\
\small Institut f\"{u}r Kernphysik, Goethe-Universit\"{a}t Frankfurt am Main\\
\small Max-von-Laue-Str. 1 Frankfurt am Main, 60438, Germany \\
\small hleon@ikf.uni-frankfurt.de
}

\maketitle

\begin{abstract}
A Monte Carlo study of two jet observables that can be used
to obtain jet samples enhanced in quark or gluon content is
presented. We compare the performance of one of these methods
on pure Monte Carlo events with that after the transport of
events through the full simulation of the ALICE experiment. It is shown that the
performance of the method is not deteriorated by experimental
effects like resolution and inefficiency.
\end{abstract}

\section{Introduction}
 
The first step towards the use of jets as probes of the properties of
the hot and dense medium produced in heavy-ion collisions is to
characterize the properties of the jets produced in vacuum. This
characterization is by itself an important test of perturbative QCD
and will also help to tune Monte Carlo generators at LHC energies.

We are interested in studying the response of the medium to quark and
gluon jets independently in order to test the theoretically predicted
differences in energy loss  \cite{1}\cite{2} due to their
different color charge factor.
The objective of this study is to test the feasibility of different
methods, in proton-proton collisions, that allow to obtain jet samples
enhanced in quark or gluon content. 

\section{ALICE charged jet reconstruction}

ALICE\cite{3}\cite{6}\cite{9} is a dedicated experiment at the LHC to study heavy-ion collisions. It has excellent tracking and particle identification capabilities\cite{3}, that makes it an ideal tool to study jet properties like fragmentation.

The jet reconstruction presented here is based only on charged particles within the ALICE acceptance $\left| \eta \right|$ $<$ 0.9. In the future, ALICE will  also be able to perform full jet reconstruction with the combination of tracking from the central barrel detectors and calorimetry with the EMCal detector.

\section{Data sample and jet finder}

The data sample used for this work consists of jet events in proton-proton collisions at $\sqrt{s}= 7$ TeV produced with Pythia\cite{4} using the set of parton distribution functions CTEQ4L\cite{5}. Full detector simulation of these events was performed using AliRoot\cite{6} in order to estimate the performance of the methods under realistic conditions.

The jet finder used was the UA1 cone algorithm, using a cone radius $R=\sqrt{\Delta\eta^{2}+\Delta\phi^{2}}$ of 0.4, a seed of 4 GeV and requiring a minimum jet transverse energy E$_{\text{T}}^{\text{Jet}}$\footnote{The UA1 jet finder algorithm produces massless jets so p$_{\text{T}}^{\text{Jet}}$$=$E$_{\text{T}}^{\text{Jet}}$.} of 5 GeV. In the following, the parton association to jets was done by selecting the most energetic parton inside a subcone with $R=$0.3 around the reconstructed jet axis. The quark/gluon content of the sample is determined this way. The charged jet reconstruction was found to introduce a bias towards finding a larger fraction of the quark initiated jets compared to the full jet reconstruction. This bias was found to increase due to the use of a small cone radius.

\section{Quark and gluon enhancement methods}

In the following subsections, the two enhancement methods studied will be described followed by a comparison of their 
performances.

\subsection{Second central moment of transverse structure}

The second central moment of the transverse structure\cite{7} is defined as in Eq.~(\ref{eq:one}).
For each component $\alpha$ ($ \alpha = \phi,\eta $) the second central moment is obtained using Eq.~(\ref{eq:two}) and  Eq.~(\ref{eq:three}), where the index $j$ corresponds to the $\eta$ and $\phi$ of the jet axis and the index $i$ runs over the charged tracks that are part of the jet, having a transverse momentum $p_{T}$.  

\begin{equation} \label{eq:one}
\left\langle \delta R^{2}_{c} \right\rangle = \left\langle \delta \phi^{2}_{c} \right\rangle + \left\langle \delta \eta^{2}_{c} \right\rangle 
\end{equation}

\begin{equation} \label{eq:two}
\left\langle \delta \alpha^{2}_{c} \right\rangle = \left\langle \delta \alpha^{2}_{j} \right\rangle - \left\langle \delta \alpha_{j} \right\rangle^{2}
\end{equation}

\begin{equation} \label{eq:three}
\left\langle \delta \alpha^{n}_{j} \right\rangle = \frac{\Sigma_{i} (\alpha_{j}-\alpha_{i})^{n} \times p^{i}_{T}}{\Sigma_{i} p^{i}_{T}} 
\end{equation}

\begin{figure}[th]
\begin{minipage}[t]{0.45\textwidth}
\begin{center}
\includegraphics[scale=0.31]{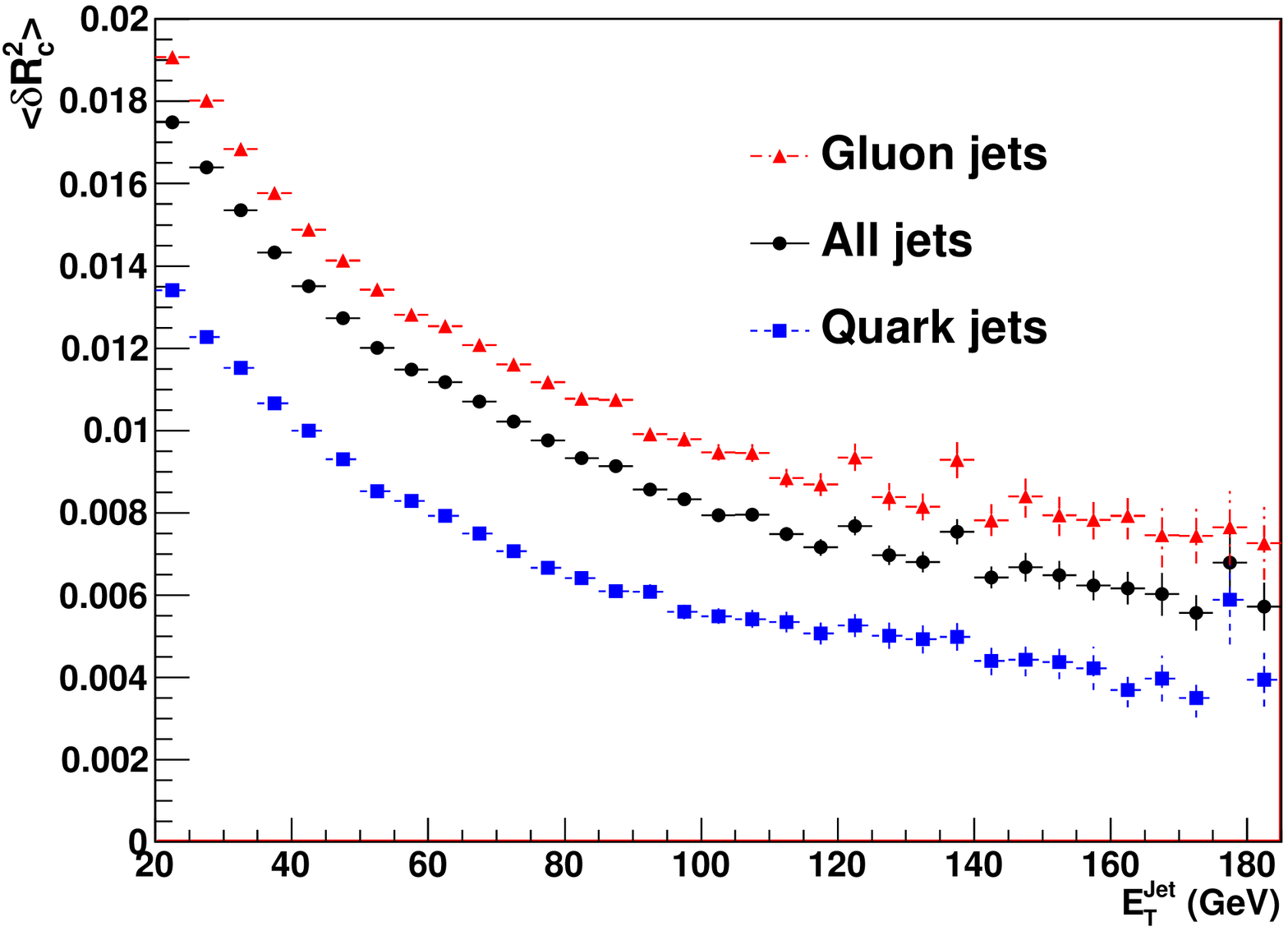}
\caption{Mean values of $\delta R_{c}^{2}$ as a function of E$_{\text{T}}^{\text{Jet}}$ for gluon jets in red triangles, quark jets in blue squares and all jets in black circles.}\label{fig:fig 2}
\end{center}
\end{minipage}
\hspace{0.5cm}
\begin{minipage}[t]{0.45\textwidth}
\begin{center}
\includegraphics[scale=0.31]{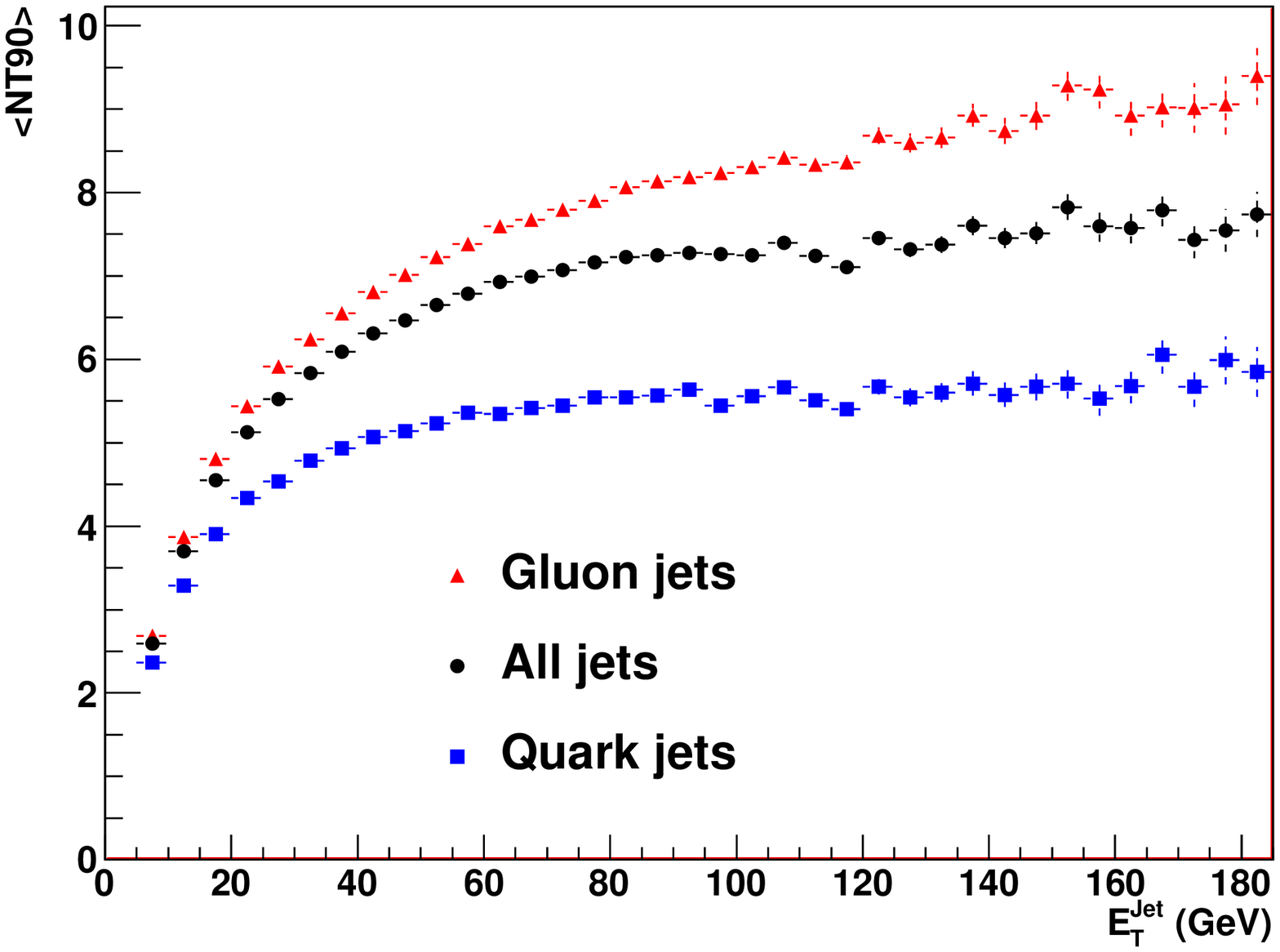}
\caption{Mean values of $NT90$ as a function of E$_{\text{T}}^{\text{Jet}}$ using the same symbols and colours as in Fig. 1.}\label{fig:fig 3}
\end{center}
\end{minipage}
\end{figure}

Figure \ref{fig:fig 2} shows the distribution of the mean values of $\delta R_{c}^{2}$ as a function of E$_{\text{T}}^{\text{Jet}}$ for quark, gluon and all jets. The tracks used to perform this analysis were charged tracks with a $p_{T} >$ 1 GeV/$c$ and with $\left| \eta \right|$ $<$ 0.9. As one can observe from the distributions, gluon jets have on average a larger mean value of $\delta R_{c}^{2}$ than quark jets over the whole range of E$_{\text{T}}^{\text{Jet}}$ studied. This difference in $\left\langle\delta R_{c}^{2}\right\rangle$ is the basis of this enhancement method.

\subsection{Track counting method}

The second method studied is based on a modified version of that proposed to be used with segmented calorimeters\cite{8}. The track counting algorithm is implemented in the following way: the tracks that compose a given jet, in this case tracks inside the cone radius around the jet axis, are ordered in decreasing magnitude of transverse momentum.  Then the transverse momentum of the ordered tracks is added until a certain fraction of E$_{\text{T}}^{\text{Jet}}$ is recovered, in this case 90\%. This minimum number of tracks is defined as the variable $NT90$.

Figure \ref{fig:fig 3} shows the distributions of the mean values of $NT90$ for quark, gluon and all jets. One can observe that the mean values of this variable are larger for gluon jets compared to quark jets of the same E$_{\text{T}}^{\text{Jet}}$. This way it is possible to select quark or gluon jets based on the multiplicity needed to recover 90\% of the E$_{\text{T}}^{\text{Jet}}$.

\subsection{Comparison of results}

For both methods, the strategy to select quark jets was based on setting an upper value on the variables so the resulting samples would be enriched in quark content, and gluon jets were selected by requiring a minimum value of $NT90$ or $\left\langle\delta R_{c}^{2}\right\rangle$.

In order to compare the performance of both methods, their results for selecting quark jets at a given jet energy and achieving the same purity of the order of 60\% to select quark jets were compared. The method of the second moment of the transverse structure is 30\% less efficient than the track counting method, as shown in Table 1.

\begin{table}[hpt] 
\caption{Comparison of discriminating methods for E$_{\text{T}}^{\text{Jet}}$ = 20 $\pm$ 5 GeV.}
\centering
{\begin{tabular}{@{}ccc@{}} 
\hline
\hphantom{000} Method \hphantom{000}   & \hphantom{000} Purity \hphantom{000} & \hphantom{000} Efficiency \hphantom{000} \\ 
\hline
$\left\langle\delta R_{c}^{2}\right\rangle$ & 61\% & 38\%  \\
$NT90$ & 64\% & 69\%  \\ 
\hline
\end{tabular}}
\end{table}

In the next step of the study, we tried to combine cuts using both variables to increase the discriminating power. The result, however, is that both variables were highly correlated. In the following, only the performance of the track counting method will be described comparing results using only Monte Carlo (MC) events and the results after the full reconstruction chain.

\section{$NT90$ method performance}

In order to evaluate the enhancement power of the variable $NT90$, the efficiency ($\varepsilon$) and purity ($P$) for  selecting quark/gluon jets was studied as a function of the cuts on $NT90$. This study was done for a fixed value of the E$_{\text{T}}^{\text{Jet}}$, the value was chosen to be 20 $\pm$ 5 GeV.

\begin{figure}[th]
\begin{minipage}[t]{0.45\textwidth}
\begin{center}
\includegraphics[scale=0.31]{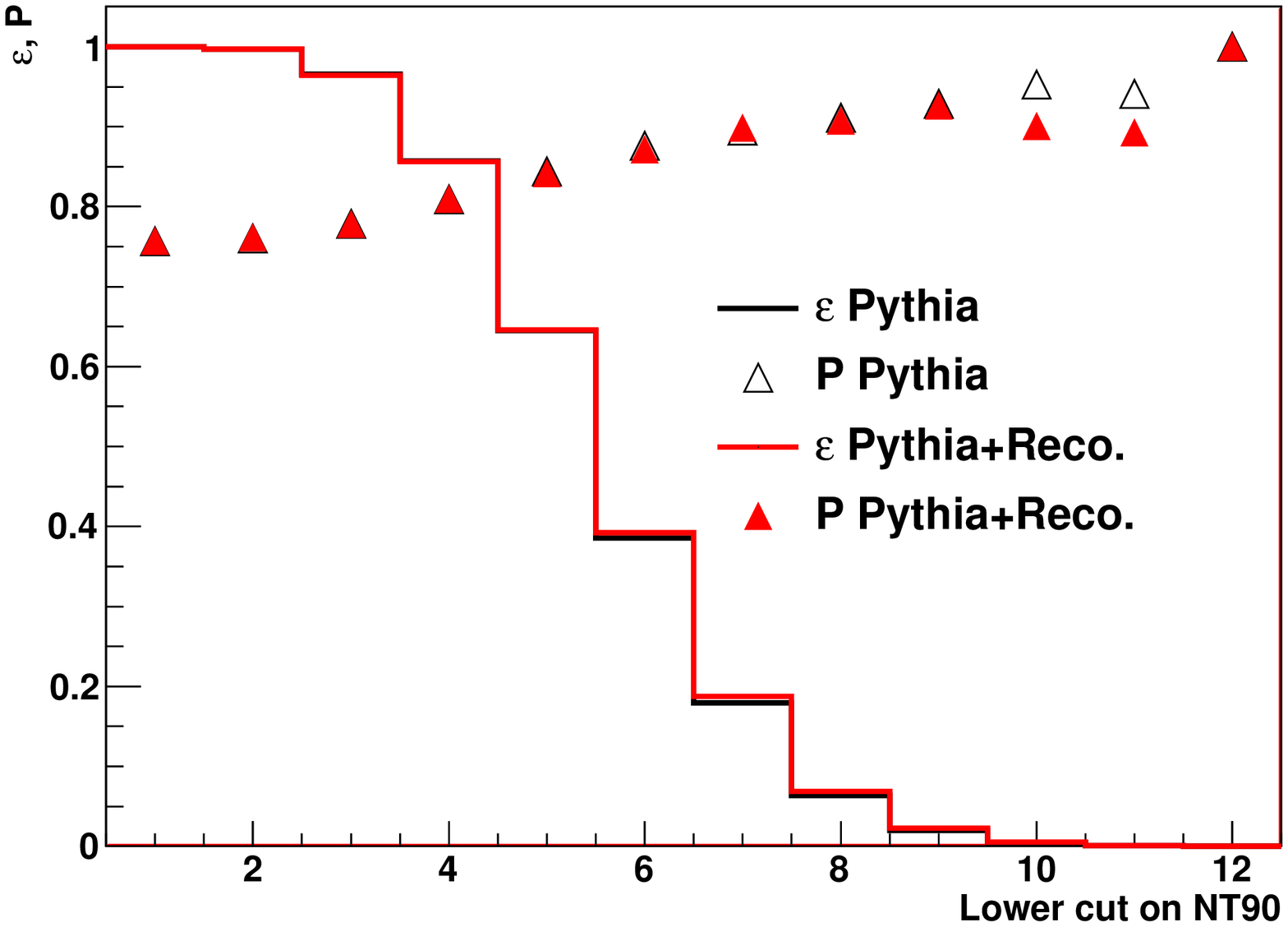}
\caption{$\varepsilon$ and $P$ for gluon jets as a function of the cut on $NT90$. The results from pure MC are shown with black lines and open triangles and the results after the ALICE simulation with a red line and full triangles.}\label{fig:fig 4}
\end{center}
\end{minipage}
\hspace{0.5cm}
\begin{minipage}[t]{0.45\textwidth}
\begin{center}
\includegraphics[scale=0.31]{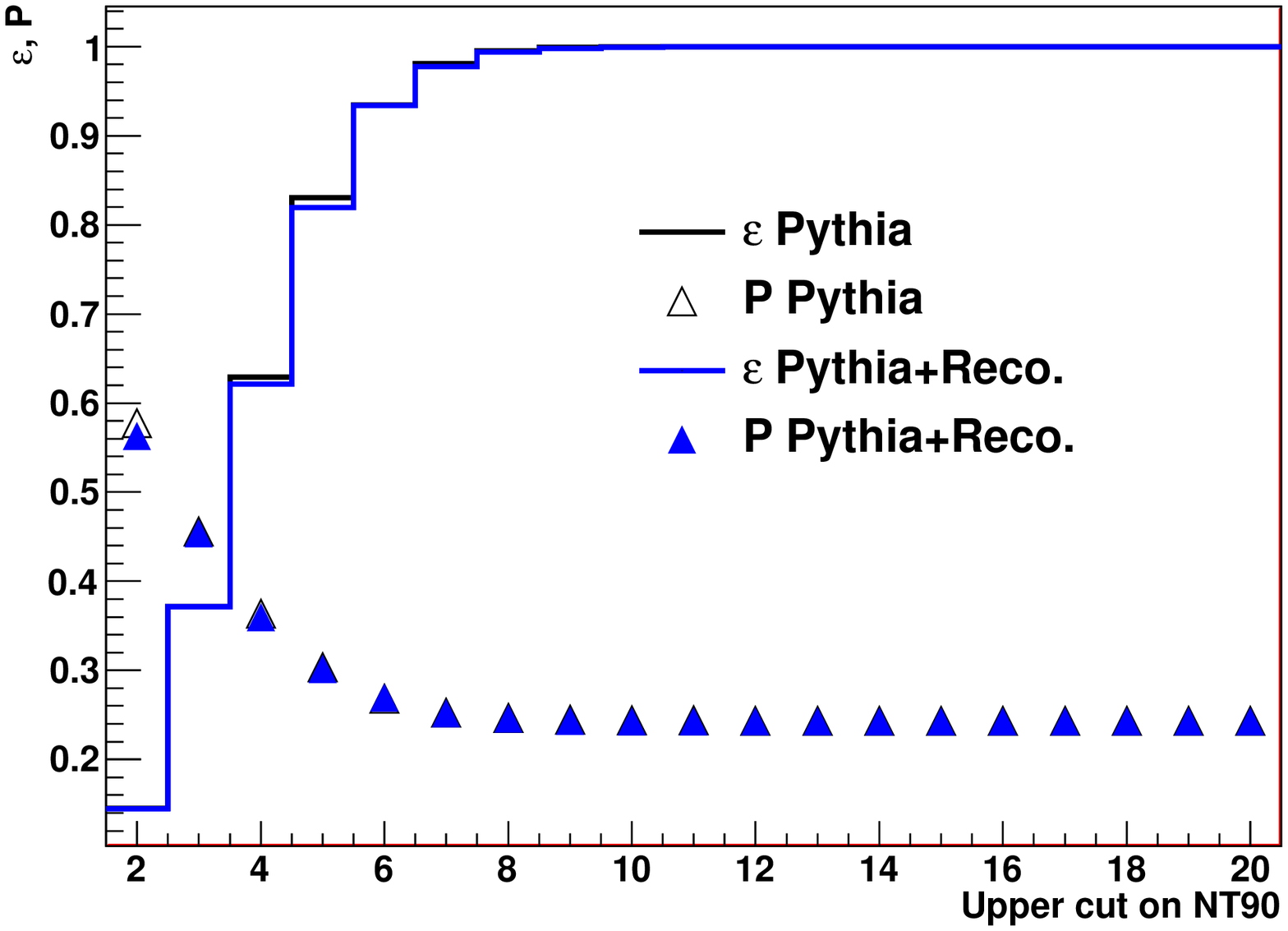}
\caption{$\varepsilon$ and $P$ for quark jets as a function of the cut on $NT90$. Results from pure MC are shown with black lines and open triangles and the results after the ALICE simulation with a blue line and full triangles.}\label{fig:fig 5}
\end{center}
\end{minipage}
\end{figure}

Figures \ref{fig:fig 4} and \ref{fig:fig 5} show the results from the pure MC events and for the same events after the full detector simulation.

For gluons jets in Figure \ref{fig:fig 4}, the first bin corresponds to not performing any cut. By setting the cuts, the minimum value on the number of tracks necessary to recover 90\% of the E$_{\text{T}}^{\text{Jet}}$ is fixed. For the quark case, shown in Figure \ref{fig:fig 5}, the last bin represents the result when no cut is applied and successive cuts are made towards the first bin, in this case setting an upper value for the value of $NT90$.

As one can see, the effect of the detector response does not affect the performance of the method. Table 2 shows an example of the purity achieved by setting cuts on $NT90$, but one should keep in mind that the method is better suited to disentangle jets produced by quark or gluon with higher E$_{\text{T}}^{\text{Jet}}$. It is in the high energy regime where the differences in the properties of quark and gluon jets are larger using $NT90$ (see Figure \ref{fig:fig 3}).

\begin{table}[hpt] 
\caption{Purity and efficiency for quark and gluon samples for E$_{\text{T}}^{\text{Jet}}$ = 20 $\pm$ 5 GeV.}
\centering
{\begin{tabular}{@{}cccc@{}} 
\hline
\hphantom{000} Parton \hphantom{000}   & \hphantom{000} Cut value \hphantom{000} & \hphantom{000} Purity (enhancement) \hphantom{000} & \hphantom{000} Efficiency \hphantom{000} \\ 
\hline
Gluon & 7 & 90\% (14\%) & 18\%  \\
Quark & 4 & 36\% (12\%) & 63\%  \\ 
\hline
\end{tabular}}
\end{table}

\section{Conclusions}

The feasibility of a method to obtain samples enriched on quark or gluon content has been studied.
It was shown using a full simulation of ALICE that it is possible to use a tagging variable based
on charged track multiplicity to obtain jet samples enhanced in quark or gluon content in proton-proton collisions. For 20 GeV 
jets it is possible to increase the purity of gluon jets up to 90\% and of quarks jets
up to 36\% having efficiencies of 18\% and 63\% respectively. Using higher E$_{\text{T}}^{\text{Jet}}$ it would be
possible to obtain samples with higher purities of quark jet content.

It is clear that the $NT90$ tagging method biases the jet samples
towards having a certain structure, i.e. soft gluon jets and hard
quark jets.  However $NT90$ can also be used as a jet structure observable that can be studied
to compare jets from proton-proton collisions with those produced from heavy-ion collisions.

\section*{Acknowledgements}

The author thanks C. Blume and C. Klein-B\"{o}sing for fruitful discussions, comments and their interest in this study.
This work was supported by the Helmholtz Research School for Quark Matter Studies.

\end{document}